\documentclass[preprint]{revtex4-1}
\usepackage{amsmath}
\usepackage{amsfonts}
\usepackage{graphicx}
\usepackage{caption}
\usepackage{bbm}

\begin{document}

\title{Time reversal symmetry and collapse models}
\date{\today}
\author{D.~J.~Bedingham}
\email{daniel.bedingham@philosophy.ox.ac.uk}
\author{O.~J.~E.~Maroney}
\email{owen.maroney@philosophy.ox.ac.uk}
\affiliation{Faculty of Philosophy, University of Oxford, OX2 6GG, United Kingdom.}

\begin{abstract}
Collapse models are modifications of quantum theory where the wave function is treated as physically real and the collapse of the wave function is a physical process.  This appears to introduce a time reversal asymmetry into the dynamics of the wave function since the collapses affect only the future state.  This paper challenges this conclusion, showing that in three different examples of time asymmetries associated with collapse models, if the physically real part of the model can be reduced to the locations in space and time about which collapses occur, then such a model works both forward and backward in time, in each case satisfying the Born rule. Despite the apparent asymmetry of the collapse process, these models in fact have time reversal symmetry.  Any physically observed time asymmetries that arise in such models are due to the asymmetric imposition of initial or final time boundary conditions, rather than from an inherent asymmetry in the dynamical law. This is the standard explanation of time asymmetric behaviour resulting from time symmetric laws.
\end{abstract}

\maketitle

\section{Introduction}

The idea of a physical wave function undergoing collapse seems to be at odds with time reversal symmetry.
For example, during the collapse process an initially dispersed wave function might become spontaneously localised about some position. By contrast, the time reverse of this process would describe an initially localised state spontaneously dispersing. Collapse models treat the collapse of the wave function as a genuine physical process and therefore inherit this time reversal asymmetry. The most well known examples are the GRW model \cite{GRW} and the CSL (continuous spontaneous localisation) model \cite{CSL1,CSL2} (for general reviews of collapse models see \cite{REP1,REP2}).
 		
There are grounds to be suspicious about the time asymmetry of wave function collapse since the Born rule can be understood to apply both forward and backward in time as outlined in the two-state vector formalism (TSVF) of Ref.\cite{ABL}. The usual way in which quantum mechanics is applied involves the construction of ensembles of pre-selected states. This allows us to make predictions about the future state. The TSVF shows that by post selecting the final state we can also make retrodictions using the same laws. This implies that the basic laws of quantum mechanics can be understood in a time symmetric way and that time asymmetry results from the way in which we choose our statistical ensembles.

In collapse models the physical collapse of the wave function is a manifestation of the Born rule. However, it is a manifestation of the Born rule only for the forward in time direction. Given that the Born rule can be stated in a time symmetric manner, we must be careful that we are not imposing our own prejudices about how the future differs from the past.

Here we will argue that collapse models can indeed be understood in a time symmetric way. In order to do this we make use of a number of ideas:
\begin{itemize}
\item
{\it Collapses happen randomly:} A typical feature of collapse models is that the wave function undergoes spontaneous collapses in some preferred basis. This process may be discrete or continuous. The probability for any given collapse to happen should be given by the Born rule. For example, in the GRW model \cite {GRW} the preferred basis is the particle position basis and the collapse process is discrete. The wave function $\psi_t(x_1,\ldots,x_N)$ for $N$ distinguishable particles usually satisfies the Schr\"odinger equation but from time to time, and with fixed probability per unit time for each particle, it makes a jump of the form
\begin{align}
\psi_{t+} = j(z - x_i)\psi_t.
\end{align}
Each particle has its own random sequence of jumps occurring at a different set of random times. The jump operator $j$ is of the form
\begin{align}
j(x) = \exp \left( -x^2/2a^2\right)/(\pi a^2)^{1/4},
\end{align}
with $a$ some fixed length scale. The action of $j$ is therefore to quasi project the wave function for particle $i$ about some position $z$ (the set $\{j^2(z-x_i)| z\in \mathbb{R} \}$ forms a POVM). The collapse centre $z$ is chosen randomly from a probability distribution
\begin{align}
\frac{\int dx_1\cdots dx_N |j(z-x_i)\psi_t|^2}{\int dx_1\cdots dx_N |\psi_t|^2}.
\end{align}
This is precisely the Born rule probability for a quasi projection of the form $j$. The fixed rate of collapses can be chosen such that individual particles are rarely affected, but a bulk mass with large numbers of particles suffers frequent jumps. In this way macroscopic pointers rapidly commit to definite readings.

\item
{\it A model of wave function collapse can be given an ontology in terms of the locations of the collapses:}
This is usually referred to as the flash ontology \cite{FLASH}. The wave function is an object that does not exist in ordinary position space. However, the jumps in the GRW model are localised in space and time. It therefore makes sense to treat the collapses themselves, or the locations of the collapse centres $z$, as the basis for the local beables of the theory - the mathematical counterparts to real world events \cite{BELL}. On a fine-grained scale the world appears as if composed from many discrete points. The local density of these points give a representation of the location of matter. In this picture the role of the wave function is in determining the probabilities for the various collapse locations.

In some other collapse models, notably the CSL model and the lattice model discussed below \cite{LAT1,LAT2,LAT3}, the jumps act on a quantum field variable and occur throughout space and time. The resulting collapse centres take the form of a classical stochastic field. Again we expect to recover a picture of the world from this classical stochastic field by some coarse graining procedure.

\item
{\it The wave function can be abandoned altogether \cite{noWF}:} As pointed out in Ref.\cite{LAT3}, if a collapse model is considered as a stochastic law for generating the collapse centres, then the wave function can be relegated to the initial time from which it does not need to evolve. The collapsing evolution of the wave function corresponds to an updating, conditioned on the history of realised collapses, of the rule for determining the probability of future collapses. In Ref.\cite{LAT3} it is demonstrated that, given a set of collapse data generated by some specific initial wave function, a generic wave function (subject to certain constraints on conserved quantities) will, after undergoing those same collapses, tend towards the same state as the original wave function.  This implies that even the initial wave function can be replaced by a sufficient period of collapse data.  The evolving wave function is then just a convenient calculation tool for making the theory Markovian.

\end{itemize}

On the basis of these ideas, we conjecture that, given a valid set of collapse centres, we can form a picture of a collapsing wave function evolving both forward and backward in time, and in each case, the locations of subsequent collapses will satisfy the Born rule. In this sense a collapse model can have time reversal symmetry. The forward going and backward going pictures will not be the same for the same set of collapse data. In particular the backward going wave function will be affected by collapse events in the future and not the past. Nonetheless, the collapse centres, the local beables of the theory, will be consistent between the two pictures.

To investigate this conjecture we will examine three different cases where collapse models might be thought to show distinctive time asymmetries: in Section \ref{lattice} the lattice model of collapse proposed by Dowker, Henson, and Herbauts \cite{LAT1,LAT2,LAT3}; in Section \ref{qmupl} the QMUPL (quantum mechanics with universal position localisation) model of Di\'osi \cite{DIO1}; and in Section \ref{equil} the generic tendency of collapse models to increase energy over time. In the first two cases we choose an initial condition, an initial wave function, and evolve forward in time in order to generate a set of collapses. We then show that there is an equivalent, backward in time picture of an evolving wave function undergoing the same set of collapses in the reverse order. We will show that the fixed collapses which were generated by the forward in time dynamics are distributed as though they had been generated by the backward in time collapse dynamics.  In the final case we will show how the apparent monotonic increase in energy is compatible with time symmetric dynamical laws and discuss the significance of asymmetric boundary conditions. We end with a summary in Section \ref{sum}.

\section{Lattice model}
\label{lattice}

Here we outline the lattice model of collapse proposed by Dowker \& Henson \cite{LAT1} and further investigated by Dowker \& Herbauts \cite{LAT2,LAT3}, in a slightly different presentation.  We demonstrate some pertinent features of the lattice model. In particular, given only the stochastic field of collapse data, it is not possible to observe that the state is in a superposition of different preferred basis states. To discern the presence of some matter in a region, we must coarse grain to remove background noise. It turns out the scales on which we need to coarse grain are also the scales on which a superposition state collapses. This is important for our time symmetric picture since we should not be able to distinguish superposition states which appear in one time direction but not the other.  We present the results of a statistical test, designed to demonstrate whether the collapses are distributed as though generated by the backward in time collapse dynamics.
 
\begin{figure}[h]
        \begin{center}
        	\includegraphics[width=10cm]{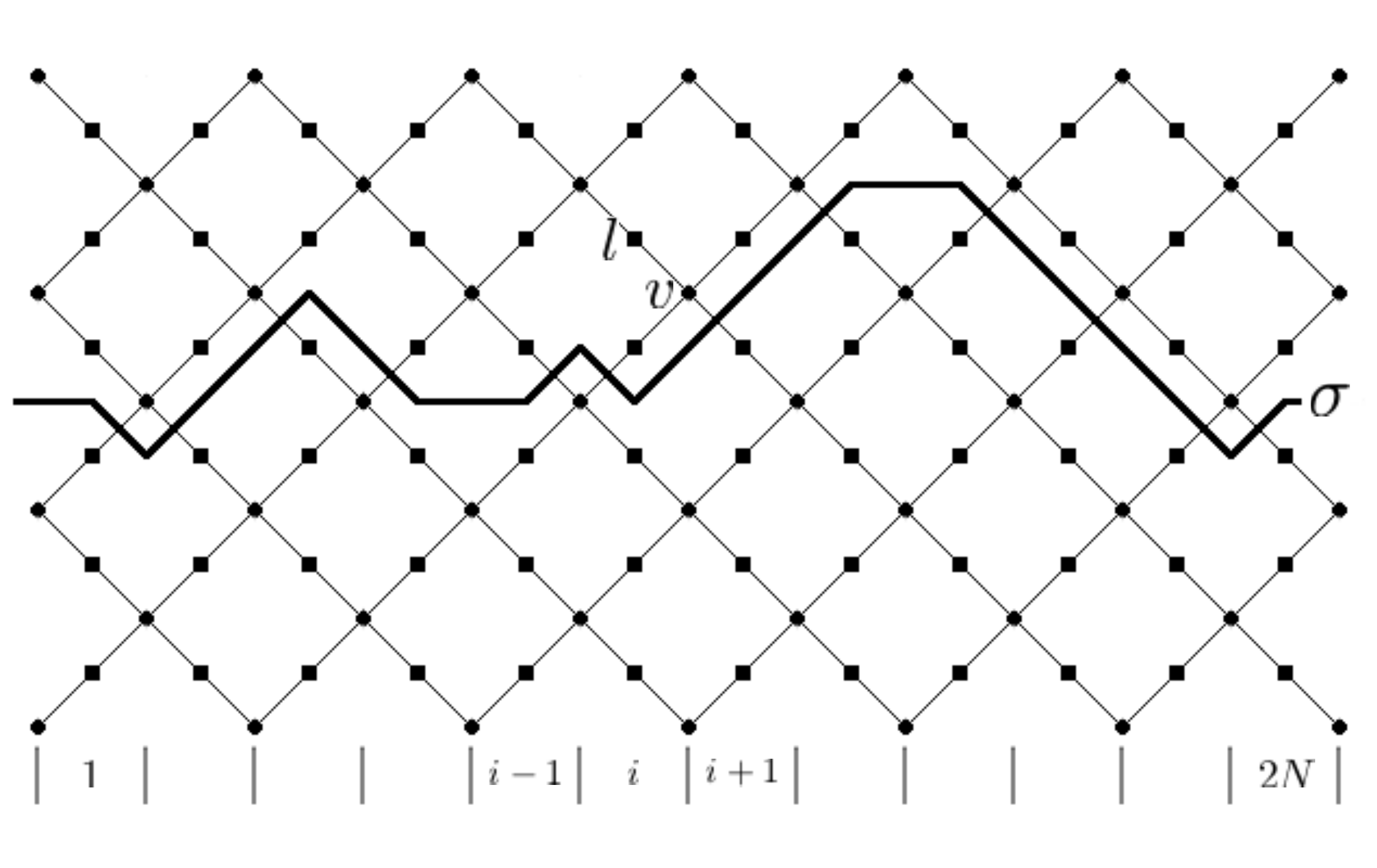}
        \end{center}
\caption{The $(1+1)$D null lattice showing a space-like surface $\sigma$, a vertex $v$, and a link $l$.}
\label{F1}
\end{figure}

 The model is a modification of the massive Thirring model on a $(1+1)$D null lattice \cite{DDV}. This is a unitary fermionic quantum field theory in 2D Minkowski space. The lattice is shown in Fig.\ref{F1}.

In the time direction (vertical) the lattice extends arbitrarily far into the past or future. In the spatial dimension there are $N$ vertices denoted by black circles and we impose periodic boundary conditions so that the space-time is in fact a cylinder. The centres of the links are also indicated by black squares. These are the locations of the collapse events. A quantum state $|\Psi_{\sigma}\rangle$ is defined with reference to a space-like surface $\sigma$. This surface must cut through the links of the lattice as shown in the figure. The surface must also satisfy the periodic boundary conditions.

We can divide the lattice into columns representing the spatial location which we label $i = 1,2,\ldots,2N$ as shown in Fig.\ref{F1}. As we pass along a space-like surface $\sigma$ we cut through each of the $2N$ columns once. To each column we assign a qubit with basis states $|0\rangle$, corresponding to empty, and $|1\rangle$, corresponding to occupied; the full quantum state is a tensor product of each of the qubit states so that the basis states of the system are of the form $|u_1, u_2, \ldots, u_{2N}\rangle$ where the $u_i  \in \{0,1\}$ represent the individual qubit basis states. For example, a state of the form $|1,0,1,1,0,0,0,1\rangle$ on a lattice of size $N=4$ can be understood either as a particular quantum field configuration or a state of $4$ `particles' at $4$ specific locations. The vacuum, or unexcited quantum field state is $|0,0,0,0,0,0,0,0\rangle$. We will refer to this basis as the preferred basis since it is the basis into which the system tends to collapse. The Hilbert space for the quantum state has dimension $2^{2N}$.

The model also makes use of a classical stochastic field $\alpha$ taking values $\alpha_{l} = 0$ or $1$ at random on each link $l$ of the lattice.

\begin{figure}[h]
        \begin{center}
        	\includegraphics[width=6cm]{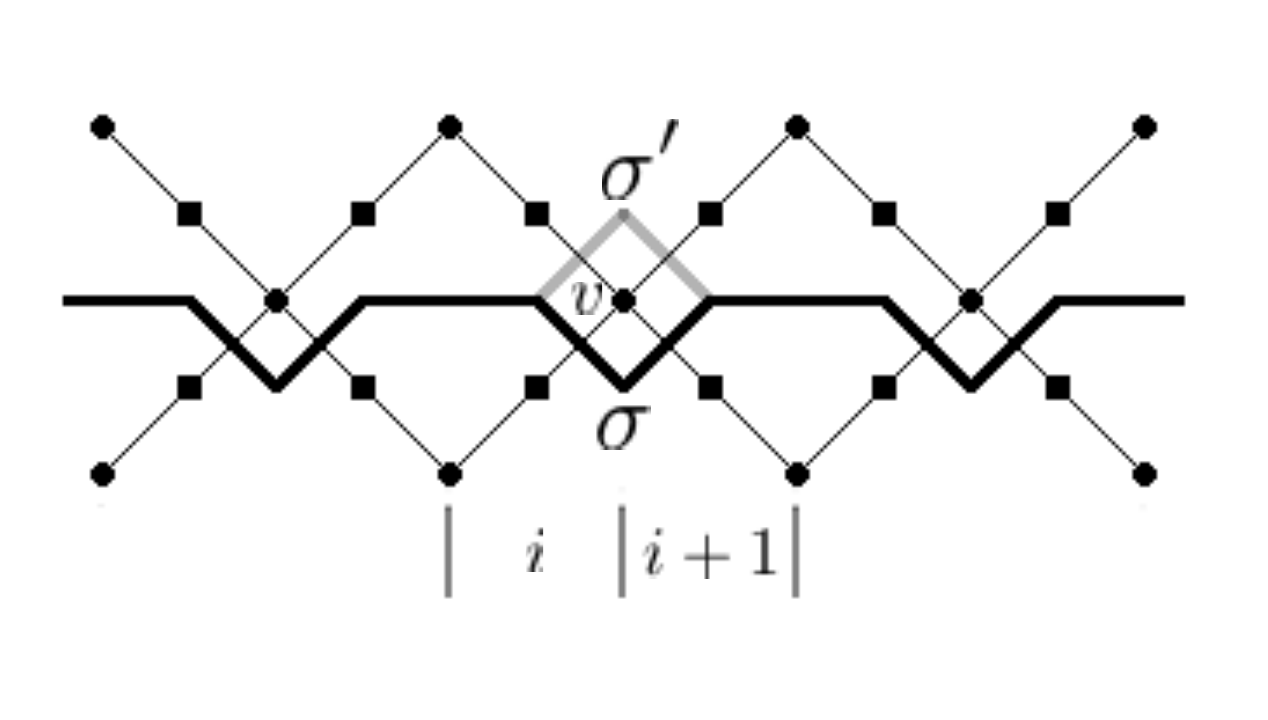}
        \end{center}
\caption{Elementary evolution I: crossing a vertex.}
\label{F2}
\end{figure}

There are two types of elementary evolution of the state which occur when the surface $\sigma$ advances in one of two different elementary ways. One corresponds the surface $\sigma$ crossing a vertex. This is shown in Fig.\ref{F2} as the surface passes over the vertex $v$ from $\sigma$ to $\sigma'$. If the vertex involved in this elementary evolution is one which connects links at positions $i$ and $i+1$ (modulo $2N$), we define a unitary operator by
\begin{align}
\mathbb{U}_v = \mathbbm{1}_1 \otimes \cdots \otimes \mathbbm{1}_{i-1}\otimes U_{i,i+1} \otimes
\mathbbm{1}_{i+2}\otimes \cdots \otimes \mathbbm{1}_{2N},
\end{align}
where $U_{i,i+1}$ is some $4$D unitary operator acting on the $i$th and $(i+1)$th qubits. In crossing the vertex $v$ the state changes according to
\begin{align}
|\Psi_{\sigma'}\rangle  =\mathbb{U}_v |\Psi_{\sigma}\rangle.
\label{evo1}
\end{align}

\begin{figure}[h]
        \begin{center}
        	\includegraphics[width=6cm]{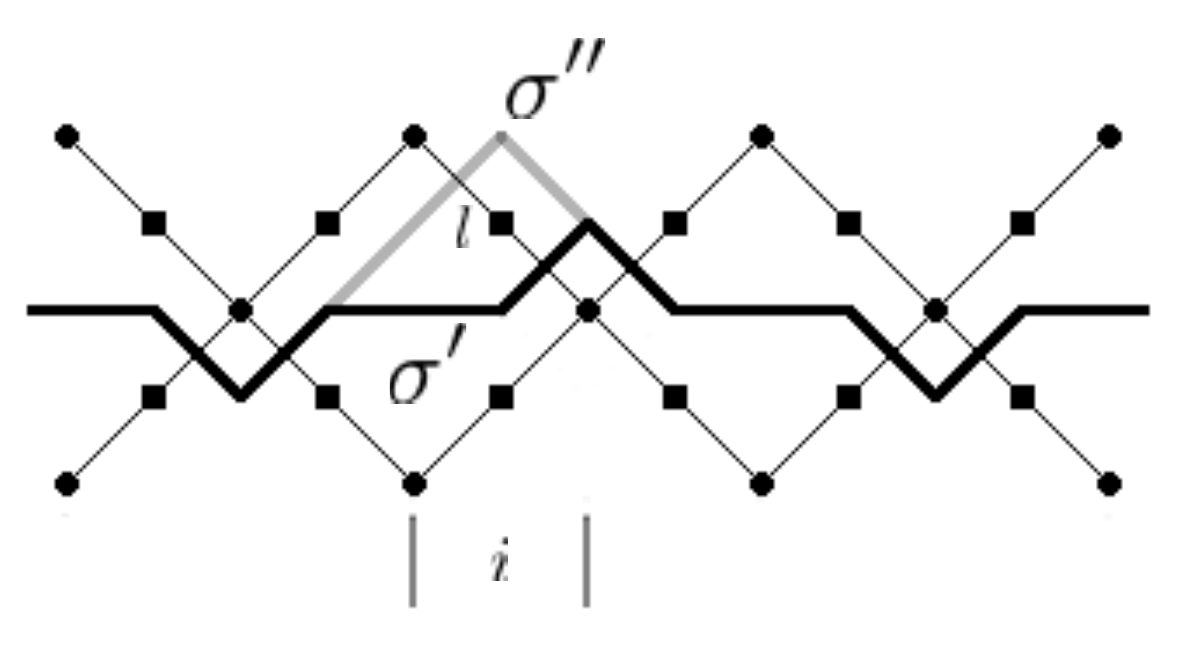}
        \end{center}
\caption{Elementary evolution II: crossing the centre of a link.}
\label{F3}
\end{figure}
The second type of elementary evolution occurs when the surface passes over the centre of a link. This is shown in Fig.\ref{F3} where the surface passes the centre of the link $l$ from $\sigma'$ to $\sigma''$. If the link is located at position $i$ on the lattice we define a jump operator by
\begin{align}
\mathbb{J}_l(\alpha_l) = \mathbbm{1}_1 \otimes \cdots \otimes \mathbbm{1}_{i-1} \otimes J_i(\alpha_{l})
\otimes \mathbbm{1}_{i+1} \otimes \cdots \otimes \mathbbm{1}_{2N}.
\end{align}
In crossing the link $l$ the state changes according to
\begin{align}
|\Psi_{\sigma''}\rangle  =
\mathbb{J}_{l}(\alpha_l)|\Psi_{\sigma'}\rangle.
\label{evo2}
\end{align}
The jump operator describes the collapse of the wave function. Without the jump operator this model is the light-cone lattice massive Thirring model of Ref.\cite{DDV}. Depending on the value of $\alpha_{l}$, the operator $J_i$ acting on the $i$th qubit takes the form
\begin{align}
J_i(0) &= \frac{1}{\sqrt{1+X^2}}
\left\{|0_i\rangle\langle 0_i| + X|1_i\rangle\langle 1_i| \right\}
\nonumber \\
J_i(1) &= \frac{1}{\sqrt{1+X^2}}
\left\{X|0_i\rangle\langle 0_i| + |1_i\rangle\langle 1_i| \right\}
\end{align}
with $X$ the fixed collapse parameter, $0 \leq X \leq 1$. The jump operators satisfy
\begin{align}
J_i^2(0) + J_i^2(1) = \mathbbm{1}_i.
\label{EQPOVM}
\end{align}
These are Krauss operators. They form a POVM on the $i$th qubit state. If $X$ is close to $0$ then whenever they act they effectively perform a projective measurement on the qubit state in the $0,1$ basis. If $X$ is close to $1$ their action nudges the qubit state slightly towards either the $0$ or $1$ state with collapse requiring many such jump operations.

The value of the field variable $\alpha_{l}$ controls whether the collapse favours the $|0\rangle$ state or the $|1\rangle$ state. This field variable is chosen randomly and the probability that the field takes value $\alpha_{l}$ on the link $l$ is given by
\begin{align}
\mathbb{P}(\alpha_{l}|\sigma')
= \frac{ \langle \Psi_{\sigma'}|\mathbb{J}^2_{l}(\alpha_l)  |\Psi_{\sigma'}\rangle}
{\langle \Psi_{\sigma'} |\Psi_{\sigma'}\rangle}
=\frac{ \langle \Psi_{\sigma''} |\Psi_{\sigma''}\rangle}
{\langle \Psi_{\sigma'} |\Psi_{\sigma'}\rangle }.
\label{EQPR}
\end{align}
This rule ensures that the jump operators act in accordance with their Born rule probabilities. This is because the probability (\ref{EQPR}) is precisely the Born probability for the quasi projection $J_i$.

From these elementary rules we can derive the rules for evolution from a general state on an initial surface $\sigma_i$ to a final surface $\sigma_f$ in the future. We suppose that in getting from $\sigma_i$ to $\sigma_f$ we must cross $n$ links and $m$ vertices. The links we label $\{l_1,\ldots,l_n \}$ and the vertices we label $\{v_1,\ldots,v_m \}$. It follows that the final state is of the form
\begin{align}
|\Psi_{\sigma_f} \rangle =
{\cal T}\left[\mathbb{J}_{l_1}(\alpha_{l_1})\cdots\mathbb{J}_{l_n}(\alpha_{l_n})\mathbb{U}_{v_1} \cdots \mathbb{U}_{v_m} \right]|\Psi_{\sigma_i}\rangle,
\label{evonm}
\end{align}
where ${\cal T}$ is the time ordering operator. The probability for the field values $\{\alpha_{l_1},\ldots,\alpha_{l_n} \}$ is given from (\ref{EQPR}) by
\begin{align}
\mathbb{P}(\alpha_{l_1},\ldots,\alpha_{l_n}|\sigma_i)
= \frac{\langle \Psi_{\sigma_f}|\Psi_{\sigma_f}\rangle}{\langle \Psi_{\sigma_i}|\Psi_{\sigma_i}\rangle}.
\label{probnm}
\end{align}
This probability is well defined given only the partial ordering of links and vertices imposed by the space-time causal order \cite{LAT1}. 

We shall work with a unitary operator of the form \cite{DDV}
\begin{align}
U_{i,i+1} = \begin{array}{cc}
		&
		\left.		
			\begin{array}{cccc}
				 \cdot \;\; \cdot \;\;\;\;\; & \cdot \nwarrow \;\;\;\; & \nearrow \cdot \;\;  & \nearrow \nwarrow
			\end{array}
		\right. \\
		\begin{array}{c}
			\cdot \;\;\cdot \\ \cdot\nearrow  \\ \nwarrow\cdot \\ \nwarrow \nearrow
		\end{array} &
		\left(
			\begin{array}{cccc}
				\;\;\; 1 \;\;\; &  0  &  0  &  \;\;\; 0 \;\;\;  \\
				0 & \; i\sin\theta \; & \; \cos\theta \; & 0 \\
				0 & \cos\theta & i\sin\theta & 0 \\
				0 & 0 & 0 & 1
			\end{array}
		\right),
	\end{array}
\end{align}
where $\nearrow \nwarrow$ denotes an incoming state where both the $i$th and the $(i+1)$th qubits are in the $|1\rangle$ state; $\nearrow\cdot$ denotes an incoming state where the $i$th qubit is in the $|1\rangle$ state and the $(i+1)$th qubit is in the $|0\rangle$ state; etc. Here, $\theta$ controls the speed of the `particles' on the lattice with, e.g., $\theta = 0$ corresponding to light speed particles and $\theta = \pi/2$ corresponding to stationary particles. All excitations of the quantum field travel with the same speed although they may move either to the left or the right. This unitary operator conserves the number of particles (the number of occupied qubits).

If we eliminate the jumps from the dynamics we can unwind the state evolution backward in time using the inverse unitary matrices
\begin{align}
|\Psi_{\sigma_i} \rangle =
\overline{\cal T}\left[\mathbb{U}^{-1}_{v_1} \cdots \mathbb{U}^{-1}_{v_m} \right]|\Psi_{\sigma_f}\rangle,
\end{align}
where $\overline{\cal T}$ is the anti time ordering operator. Taking the complex conjugate gives
\begin{align}
|\Psi^{*}_{\sigma_i} \rangle =
\overline{\cal T}\left[\mathbb{U}_{v_1} \cdots \mathbb{U}_{v_m} \right]|\Psi^{*}_{\sigma_f}\rangle,
\end{align}
demonstrating that the complex conjugate state obeys the usual unitary dynamics backward in time.

Now we include the jumps and suppose that the backward in time dynamics satisfy
\begin{align}
|\Phi^*_{\sigma_i} \rangle =
\overline{\cal T}\left[\mathbb{J}_{l_1}(\alpha_{l_1})\cdots\mathbb{J}_{l_n}(\alpha_{l_n})\mathbb{U}_{v_1} \cdots \mathbb{U}_{v_m} \right]|\Psi^*_{\sigma_f}\rangle.
\label{evorev}
\end{align}
This looks the same as the forward in time rule (\ref{evonm}) other than that the time ordering has been replaced by anti time ordering. In general $|\Phi_{\sigma_i} \rangle$ will not be the same as $|\Psi_{\sigma_i} \rangle$. Our aim is to show that the field values $\{\alpha_{l_1},\ldots,\alpha_{l_n} \}$ (which are fixed by the forward evolution) also satisfy the probability rule
\begin{align}
\mathbb{P}(\alpha_{l_1},\ldots,\alpha_{l_n}|\sigma_f)
= \frac{\langle \Phi^*_{\sigma_i}|\Phi^*_{\sigma_i}\rangle}{\langle \Psi^*_{\sigma_f}|\Psi^*_{\sigma_f}\rangle},
\end{align}
the counterpart to Eq.(\ref{probnm}). If this is the case then the backward in time evolution uses precisely the same dynamical rule as the forward in time evolution. The wave functions for the two cases will differ in general on any given surface $\sigma$ but the stochastic field $\alpha$, the basis for local beables, will be consistent.

\subsection{Coarse graining and collapse time scale}
\label{coarse}

Given a set of collapses (the field $\alpha$ in our lattice model), the forward in time wave function undergoing these collapses is likely to look (at least at the micro level) different from the backward in time wave function. If the time symmetric picture of physical collapse is to work, it must be the case that we are not able to observe the state of the wave function from the collapse data in a way which would allow us to distinguish these two cases.

In this section we adapt a calculation given in Ref.\cite{LAT2} and show that the time scale necessary to observe the presence of an excited quantum field on the lattice is the same time scale on which a superposition of such excited states will collapse. This means that it is not possible to directly observe a superposition in the preferred basis given only $\alpha$.

Suppose that the system is in the vacuum state $|0,0,\cdots, 0\rangle$. The evolution of the system defined by Eqs (\ref{evo1}) and (\ref{evo2}) will not change the state. The stochastic field $\alpha$ will take value $0$ with probability $1/(1+X^2)$ and $1$ with probability $X^2/(1+X^2)$. We assume that $X$ has a value close to $1$. This is natural since we do not want single particle states to rapidly collapse. Consider a region of space-time $R$ containing $M$ links. Within this region the mean and variance of the field value are given by the binomial distribution
\begin{align}
\mu = \frac{X^2}{1+X^2} \; ,\; \sigma^2 = \frac{X^2}{M(1+X^2)^2}.
\end{align}
Now suppose that there is some non-vacuum state with mean field value $\alpha_R$ in $R$. If we are to be able to observe $\alpha_R$ against the background noise we require
\begin{align}
\sigma \ll |\alpha_R - \mu | .
\end{align}
If we write $\epsilon = 1 - X$, taking $\epsilon$ to be small, we find that $ |\alpha_R - \mu |$ is at most ${\cal O}(\epsilon)$ (occurring when the state is maximally excited in the region $R$). Since $\sigma \sim M^{-1/2}$ we therefore must have
\begin{align}
M \gg \epsilon^{-2}.
\label{EQCG}
\end{align}
This determines the size of the region necessary to be able to identify an excited state.

Now consider a block of $n$ qubits in the $1$ state with all other qubits in the $0$ state, superposed with a disjoint block of $n$ qubits in the $1$ state with all other qubits in the $0$ state. For simplicity we assume that $\theta = \pi/2$ so that the unitary evolution has no effect. Write the state as $|A\rangle + |B\rangle$.

After we have evolved for a number $m$ of time units the state will be of the unnormalised form $X^{M_B}|A\rangle + X^{M_A}|B\rangle$, where $M_A + M_B = 2nm = M$, and $M_{A/B}$ is the number of links for which the field $\alpha_l$ corresponds to the $A/B$ state qubit eigenvalue on that link wherever the $A$ and $B$ state qubit eigenvalues are different.

Since the value of $\epsilon$ is small, the field takes values $0$ or $1$ with probability $\sim 1/2$. The division of $M$ into $M_A$ and $M_B$ has a distribution with standard deviation in $M_{A/B} \propto \sqrt{M}$. This means that one of the states will be suppressed with respect to the other by a factor of $X^{\sqrt{M}} \simeq \exp (-\epsilon \sqrt{M})$. Therefore there is exponential suppression when
\begin{align}
M \sim \epsilon^{-2}.
\end{align}
We now put this together with result (\ref{EQCG}). If the ontology of the lattice collapse model is given by the stochastic field $\alpha$ then any feature observable against background noise requires coarse graining over a region containing $\gg \epsilon^{-2}$ links. Since a superposition state is only able to survive for $\sim \epsilon^{-2}$ links then we cannot directly observe the fluctuating densities during the collapse process.

Consider the example above and suppose that the state collapses to the $|B\rangle$ state after a certain period of time. If we then evolve the state backward in time it will simply stay in the $|B\rangle$ state. This can be consistent since the period during which the state was in a superposition of $|A\rangle$ and $|B\rangle$ when viewed forward in time is brief enough that it cannot be distinguished from $|B\rangle$ if we only have access to the stochastic field $\alpha$.

\subsection{Time reversed collapse on the lattice}
\label{latticetest}

To test for consistency between forward in time and backward in time collapse dynamics on the lattice we apply the following test:
\begin{enumerate}
\item We generate a some field data by starting with some initial state $|\Psi_0\rangle$ on an initial time slice $t=0$ and evolving forward in time. This results in a random field $\alpha$.

\item We then reverse the evolution in time using Eq.(\ref{evorev}) applying the collapses again in the reverse order using the same field data $\alpha$. For each link $l$ we make note of the probability that the field $\alpha_l$ takes the value $1$ conditional on all $\alpha$ to the future of $l$. These probabilities can take values anywhere in the range $[0,1]$. Once we have returned to time $t=0$ we have a set of reverse time probabilities for the field on each link on the lattice to take value $1$ alongside the realised set of field values $\alpha$.

\item We divide the set of reverse time probabilities into a set of bins with boundaries $0<p_1<p_2<\ldots<1$. For each bin $j$ we take the average probability as $\bar{p}_j = (p_{j-1} - p_j)/2$ and we count the number of collapse events in that bin, $m_j$. Using the binomial distribution we determine the mean number of times that we expect to see the stochastic field take value 1, $\mu_j$, and the variance in this number, $\sigma_j^2$. We use a simple test to check that the data should be approximately normally distributed (essentially that $m_j$ should be sufficiently large given the average probability $\bar{p}_j$ for the bin) and discard those bins where this is not the case. We then count the actual number of times $n_j$ that the field $\alpha$ for each of those events realised the value $1$. From these we calculate a chi-squared statistic
\begin{align}
\chi^2 = \sum_{j}\frac{(n_j - \mu_j)^2}{\sigma_j^2}.
\label{chi}
\end{align}

\item Using the theoretical distribution for the chi-squared statistic we calculate a p-value giving the probability that the field data or something more extreme could have been generated by the reverse time probabilities. In statistics a p-value of less than 0.5\% is a typical standard for ruling out a hypothetical model.

\end{enumerate}

\begin{figure}[h]
        \begin{center}
        	\includegraphics[width=16cm]{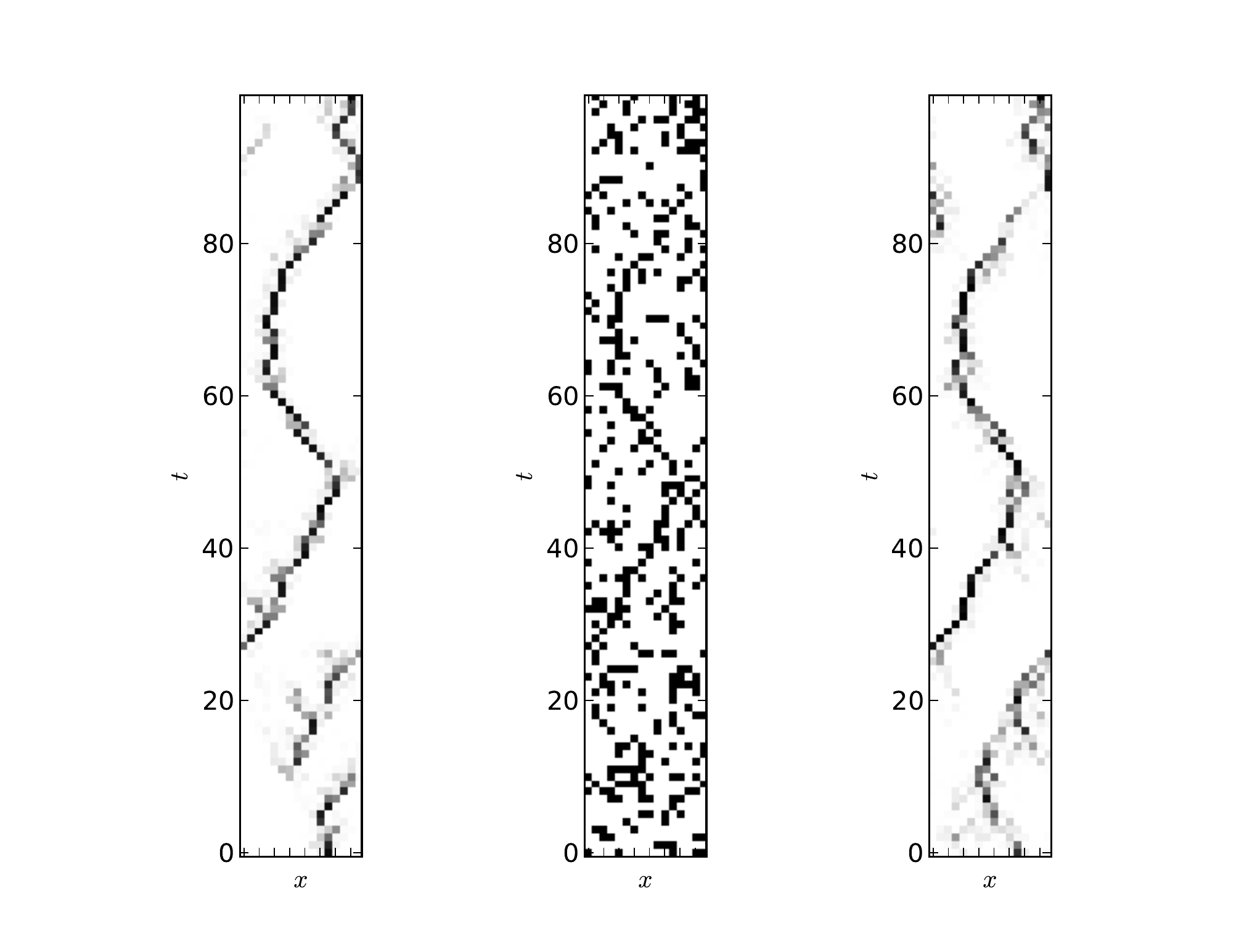}
        \end{center}
\vspace{-1.2cm}
\hspace{0.5cm}(a) \hspace{3.6cm} (b) \hspace{3.6cm} (c)
\caption{(a) Quantum field expectation value for the forward in time state. (b) Stochastic field $\alpha$. (c) Quantum field expectation value for the backward in time state.}
\label{F4}
\end{figure}

Figure \ref{F4} shows a typical example of a simulation with $2N = 16$ in which we set $X=0.5$ and $\theta = \pi/4$. The initial condition is of the form
\begin{align}
|\Psi_0\rangle = |0,0,0,0,0,0,0,0,0,0,1,0,0,0,0,0\rangle.
\end{align}
This is a single `particle' state for the quantum field. We evolve for $100$ time steps. We choose an arbitrary total ordering of links and vertices such that elementary evolutions sweep from left to right. (To be precise we start on a given time slice, we then evolve past the leftmost vertex, followed by the link above it to the left and then the link above it to the right. We then do the same for the vertex to the right and so on until we have evolved the time slice by one time step. The procedure is then repeated.) Each pixel in the figure represents a link on the lattice so that each graph is $16$ pixels wide and 100 pixels high.

The left hand panel shows the expectation value $\langle\Psi_\sigma |\mathbb{A}_l|\Psi_\sigma\rangle$ where
\begin{align}
\mathbb{A}_l = \mathbbm{1}_1 \otimes \cdots \otimes \mathbbm{1}_{i-1} \otimes A_i
\otimes \mathbbm{1}_{i+1} \otimes \cdots \otimes \mathbbm{1}_{2N}.
\end{align}
and
\begin{align}
A_i =
|1_i\rangle\langle 1_i|.
\end{align}
This is the expectation of the eigenvalue of the $i$th qubit. The surface $\sigma$ is the surface immediately following evolution across the link $l$ given the total ordering which we use. We use a grey scale to plot the values of $\langle\Psi_\sigma |\mathbb{A}_l|\Psi_\sigma\rangle$ with black corresponding to the value $1$ and white corresponding to $0$. A clear particle trajectory winding around the lattice is apparent.

The central panel shows the realised field values $\alpha$ on the lattice generated randomly using the probabilities calculated from the forward evolving state $|\Psi_{\sigma}\rangle$. Here black corresponds to $\alpha_l = 1$ and white corresponds to $\alpha_l = 0$. Here the particle trajectory is less apparent. As discussed in Sec.\ref{coarse} a coarse graining procedure is necessary to eliminate background noise and establish the particle's whereabouts to within space time regions of greater than $1$ pixel.

The right hand panel again shows $\langle\Psi_\sigma |\mathbb{A}_l|\Psi_\sigma\rangle$, but this time the state evolves backward in time using the reversed total ordering and with collapses generated by the same field $\alpha$. The particle trajectory is very similar to that in panel (a).

A careful observation of Fig.{\ref{F4}}(a) reveals that dispersion occurs in the upward direction: localised particle states tend to become diffuse with increase in $t$ before recollapsing with dead ends fading away. The same happens in the opposite time direction in Fig.{\ref{F4}}(c). Despite these differences the overall picture of a particle moving about on the lattice is broadly consistent.

\begin{figure}[h]
        \begin{center}
        	\includegraphics[width=12cm]{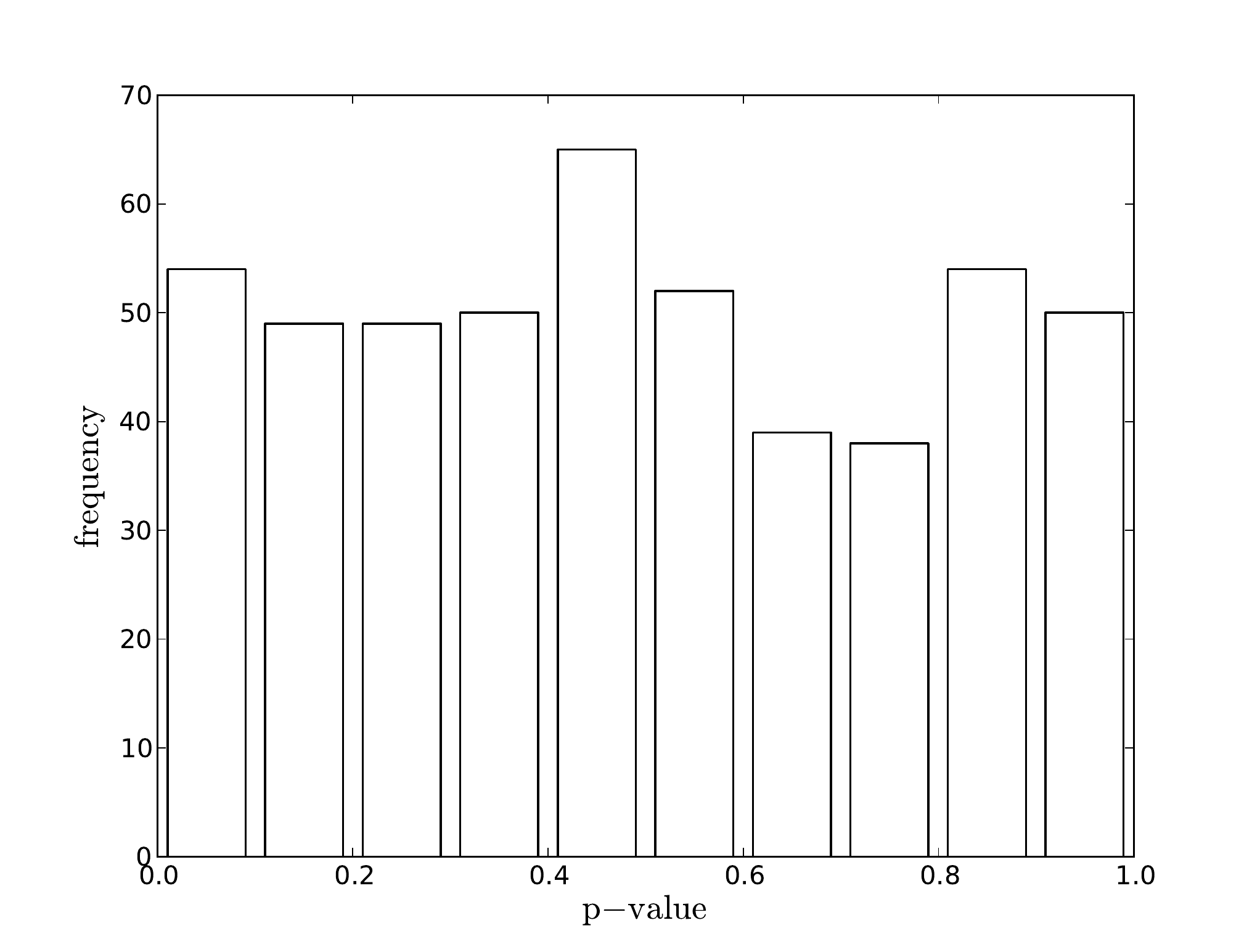}
        \end{center}
\caption{Distribution of p-values for 500 simulations}
\label{F5}
\end{figure}

As outlined above we calculate the chi-squared statistic (\ref{chi}) and a p-value for the field data in panel (b) under the null hypothesis that data was generated by the backward in time collapse dynamics. In Fig.\ref{F5} we plot a histogram of the p-values generated by running the simulation $500$ times. The distribution of p-values is approximately uniform over the range $[0,1]$. This is as we would expect if the null hypothesis is correct.

We conclude that field data generated by forward in time collapse dynamics is distributed as if it had been generated by an equivalent backward in time collapse dynamics. In this sense the collapse dynamics satisfies a form of time reversal symmetry. This suggests that a backward in time moving observer could use the same mathematical procedure, based on a collapsing wave function to determine the past state of the stochastic field, as a forward in time moving observer would use to estimate the future state of the field.

\section{Time reversal symmetry for a localised wave packet}
\label{qmupl}

We now turn our attention to the QMUPL model \cite{DIO1} for a particle in a localised state.  When the particle is localised the collapse dynamics reduce to a classical diffusion process. We demonstrate that this process is asymmetric with respect to time reversal, but show it is still possible to construct an equivalent, backward in time process resulting from the same collapses occurring in the reverse order.

The QMUPL model can be thought of as a continuous version of GRW describing wave function collapse for distinguishable particles. It is also a limit of the CSL (continuous spontaneous localisation) model when the particle density is low and the collapse length scale is large compared to the length scale of a wave packet \cite{ME1}. In the QMUPL model the state vector for a single particle satisfies a quantum state diffusion of the form
\begin{align}
	d|\psi_t\rangle = \left\{-i\hat{H} dt  - \frac{g^2}{8}( \hat{x}-\langle \hat{x}\rangle_t)^2 dt
+ \frac{g}{2}(\hat{x}-\langle \hat{x} \rangle_t) d{B}_{t}\right\}|\psi_t\rangle.
\label{QSD}
\end{align}
Here $B_t$ is a standard Brownian motion and the collapse parameter $g$ controls the rate at which collapse of the wave function occurs.

A typical feature of continuous collapse models is that after a sufficient period of time, the wave packet of an individual isolated particle achieves a stable localised  shape. This happens when the dispersive effects of quantum theory balance with the localising effects of the collapses. When the particle is in this condition the dynamics are considerably simplified. The state can be characterised by only the central values of position $\langle \hat{x} \rangle_t = \langle \psi_t | \hat{x} |\psi_t \rangle = x_t$ and momentum $\langle \hat{p} \rangle_t =\langle \psi_t | \hat{p} |\psi_t \rangle= p_t$ for the wave packet. The collapse dynamics causes these phase space parameters to undergo a classical diffusion process. For the QMUPL model this process is given by \cite{DIO2,DUR,PEA}
\begin{align}
	dx_t &= \frac{p_t}{m}dt
		+\frac{1}{\sqrt{m}} d B_{t} ,
	\label{x} \\
	dp_t &=  \frac{g}{2} d B_{t} .
	\label{p}
\end{align}
These equations describe a classical diffusion through phase space. In what follows we restrict our attention to this simplified model of a diffusing wave packet. This avoids much of the complication of an arbitrary wave function undergoing collapse.

We first show that Eqs (\ref{x}) and (\ref{p}) do not have time reversal symmetry. The state of the system at any point in time is described by $x$, $p$, and $B$. Consider a sequence of two states $S_1$ and $S_2$ at times $t_1 =t$ and $t_2 = t+\Delta t$ respectively. We write
\begin{align}
S_1 &= \{x, p, B\} \\
S_2 &= \{x+\Delta x, p +\Delta p,  B + \Delta B\}
\end{align}
Consider the change as we go from state $S_1$ to $S_2$. The change in $x$ is $\Delta x$, the change in $p$ is $\Delta p$ and the change in $B$ is $\Delta B$. From Eqs (\ref{x}) and (\ref{p}) we expect that
\begin{align}
	\Delta x &= \frac{p}{m}\Delta t
		+\frac{1}{\sqrt{m}} \Delta B ,
	\label{x2}\\
	\Delta p &=  \frac{g}{2} \Delta B.
	\label{p2}
\end{align}
We next define a time reversal transformation $T$. For a classical phase space trajectory this involves a change in the sign of $p$ simply because playing the movie of these events backward in time, the particle appears to move in the opposite direction to when the movie is played forward in time. The time reversed states are therefore given by
\begin{align}
S_1^T &= \{x^T,p^T,B^T\} = \{x,-p,B\} \\
S_2^T &= \{x+\Delta x, -p -\Delta p, B + \Delta B\}
\end{align}
Consider now the change in going from $S_2^T$ to $S_1^T$. The change in $x$, $\Delta x^T=-\Delta x$, the change in $p$, $\Delta p^T = \Delta p$, and the change in $B$, $\Delta B^T = -\Delta B$. Inserting into Eqs (\ref{x2}) and (\ref{p2}) results in
\begin{align}
\Delta x^T &= \frac{p^T}{m}\Delta t + \frac{1}{\sqrt{m}} \Delta B^T  \\
\Delta  p^T &=   -\frac{g}{2} \Delta B^T .
\end{align}
This is a different pair of equations from those which describe evolution forward in time.  This would appear to mean that it would be possible to determine the forward direction of time from a given sequence of states. For example, we could simply observe $\Delta x$ and $\Delta p$ and see if there is a positive or negative correlation in the spontaneous jumps. If the correlation is positive then the evolution is forward in time; if the correlation is negative then the evolution is backward in time.

As with the lattice model we aim to show that this is just an artefact of the mathematical formulation and that it is possible to understand what is going on in a time symmetric way. Note that the expectation values $x_t$ and $p_t$ are features of the wave function. If we regard the wave function as no more than a convenient way to encode the collapse history then it may be that we are adding in the time asymmetry by this procedure. Let us return to the idea that the collapse locations are fundamental.

The reason for the stochastic motion of $x_t$ and $p_t$ is that collapses are occurring randomly on either side of the centre of the wave packet causing it to spontaneously jump about. By taking the continuous limit of the GRW model we can show that the locations of the collapse centres are given by
\begin{align}
z_t = x_t + \frac{1}{g}\frac{\Delta B_t}{\Delta t}.
\label{qmpz}
\end{align}
The collapse centres have a white noise distribution about the expected position $x_t$.

Now suppose that we only have information about $z_t$, i.e.~the collapse locations form the basis of our local beables. We would like to confirm whether it is possible to give different but consistent pictures of the evolutions of $x_t$ and $p_t$ in either time direction, each satisfying the same dynamical law. We perform the following test analogous to the test carried out in Sec.\ref{latticetest}:
\begin{enumerate}
\item We generate some collapse centre data $z_i$, $i = 0,\ldots, n$ by starting with an initial localised state characterised by the central position $x_0$ and momentum $p_0$ of the wave packet and evolving forward in time for $n$ discrete time steps using the rule
\begin{align}
	x_{i+1} &= x_i + \frac{p_i}{m}\Delta t +\frac{1}{\sqrt{m}} \Delta B_i , \\
	p_{i+1} &= p_i + \frac{g}{2} \Delta B_i, \\
	z_i &= x_i + \frac{1}{g}\frac{\Delta B_i}{\Delta t}.
\end{align}
Brownian increments $\Delta B_i$ are generated randomly. This results in a forward in time phase space trajectory $x_i$, $p_i$.

\item Next consider the collapses from stage 1, $z_i$, in reverse order. Denote the backward in time phase space trajectory by $x'_i$, $p'_i$.  Starting at $x'_n = x_n$ and $p'_n = -p_n$ we use the dynamical law
\begin{align}
	\Delta B'_{i-1}&= g\Delta t (z_{i-1} - x'_{i}), \\
	x'_{i-1} &= x'_i + \frac{p'_i}{m}\Delta t +\frac{1}{\sqrt{m}} \Delta B'_{i-1} , \\
	p'_{i-1} &= p'_i + \frac{g}{2} \Delta B'_{i-1}.
\end{align}
In the first equation we back out the Brownian increments from the collapse centres using (\ref{qmpz}). The result is a backward in time phase space trajectory $x'_i$, $p'_i$.

\item We perform a Kolmogorov-Smirnov test on the set of implied increments $\Delta B'_i/\sqrt{\Delta t}$ to see if they fit a normal distribution. This results in p-value for the reverse time $\Delta B'$ data. This tests whether the collapses occur with white noise distribution about $x'$. If they do then the backward in time trajectory satisfies the same dynamical law based on the collapsing wave function as the forward in time trajectory.

\end{enumerate}

\begin{figure}[h]
        \begin{center}
        	\includegraphics[width=12cm]{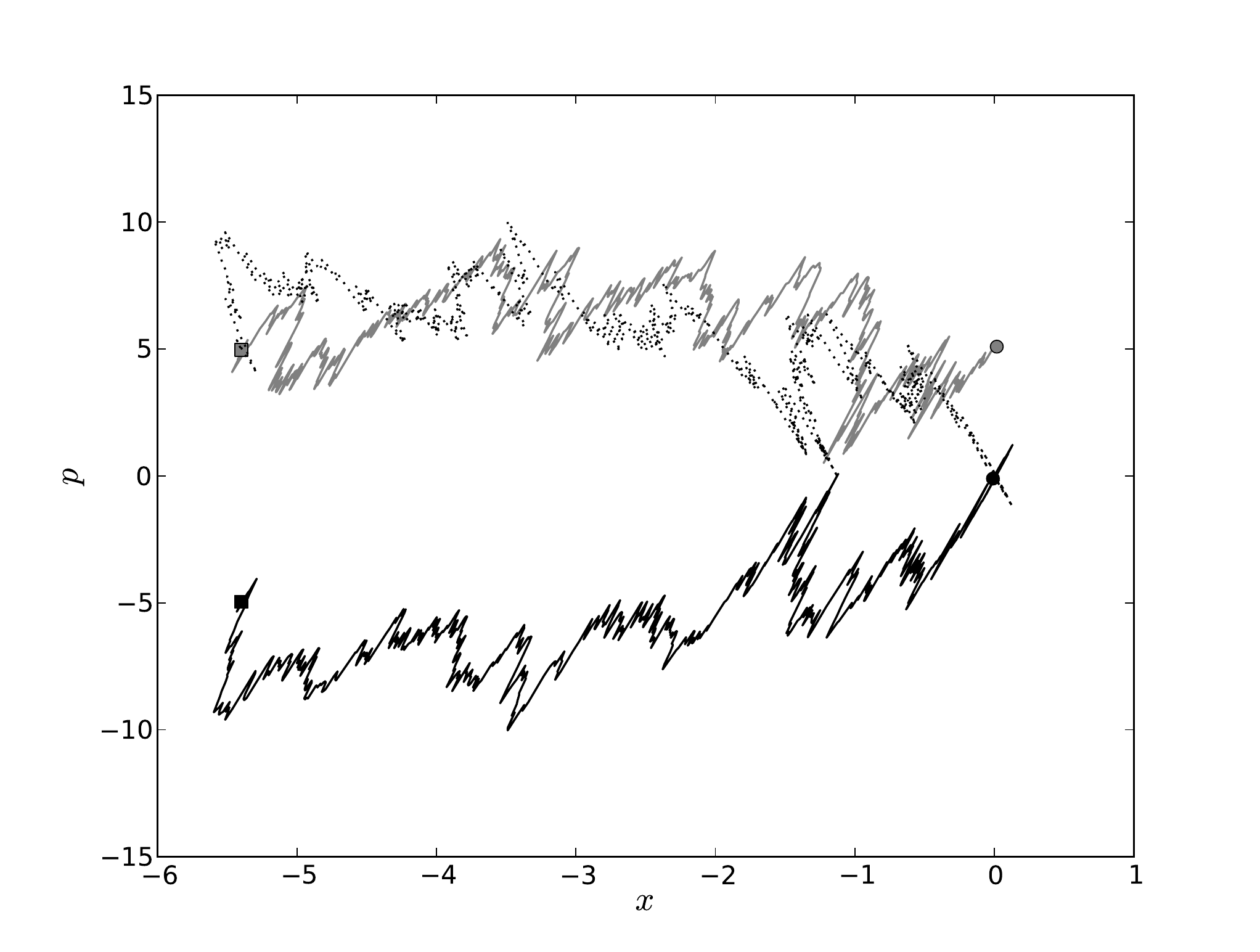}
        \end{center}
\caption{Phase space trajectories forward in time (black line), forward in time with $p\rightarrow -p$ (dotted line), and backward in time (grey line).}
\label{F6}
\end{figure}

Figure \ref{F6} shows an example of a trajectory though phase space using $g=20$, $m=1$ and $\Delta t = 0.001$. The solid black line is the forward in time trajectory generated using Eqs (\ref{x2}) and (\ref{p2}). The trajectory starts at time $t=0$ at the black circle at $x=0$, $p=0$ and ends at time $t=1$ at the black square. There is a clear positive correlation in the stochastic jumps in $x$ and $p$. The dotted line is the straightforward time reverse of this trajectory obtained by transforming $p\rightarrow -p$. It is characteristically different from the forward in time trajectory. The correlation between stochastic jumps in $x$ and $p$ is negative. The grey line shows the reverse time trajectory determined by the procedure outlined above. This trajectory starts at the grey square at time $t=1$ and ends at the grey circle at time $t=0$. This trajectory approximates the dotted line but with positively correlated jumps in $x$ and $p$. Visually, there is no way to distinguish the micro dynamics of the backward trajectory from the forward trajectory.

\begin{figure}[h]
        \begin{center}
        	\includegraphics[width=12cm]{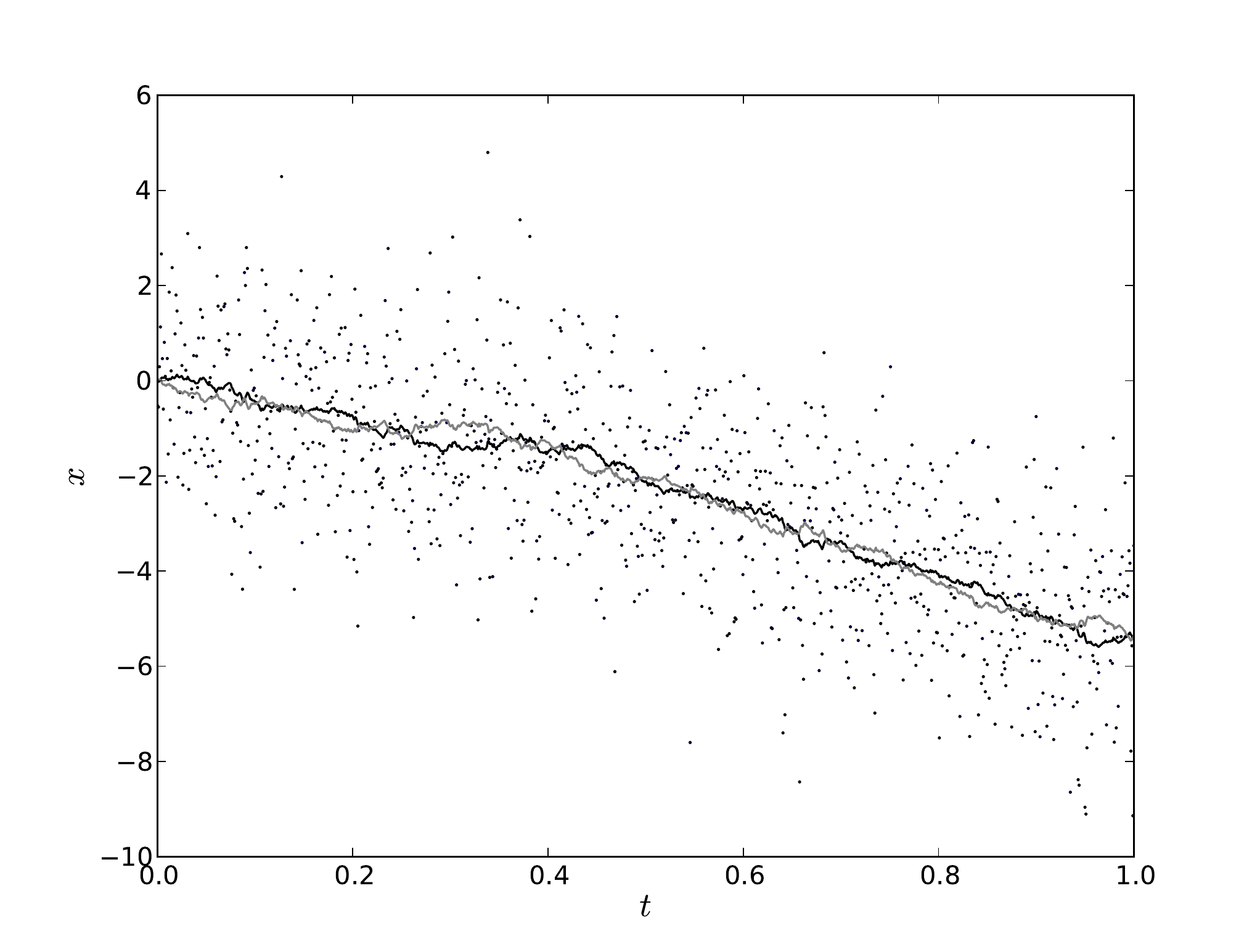}
        \end{center}
\caption{Expected position with time. Black line is evolved forward in time; grey line is evolved backward in time.}
\label{F7}
\end{figure}

Figure \ref{F7} is from the same simulated trajectory. It shows how $x$ changes with time. The dots are the collapse centres generated by the forward in time dynamics and used again in the backward in time evolution. The black line is the forward in time expectation $x_t$. It is seen that the collapse centres are distributed about this line and both the collapse centres and $x_t$ follow the same trend. The grey line is the backward in time $x'_t$ which is slightly different from the forward in time $x_t$ but which also sits well within the distribution of collapses.

\begin{figure}[h]
        \begin{center}
        	\includegraphics[width=12cm]{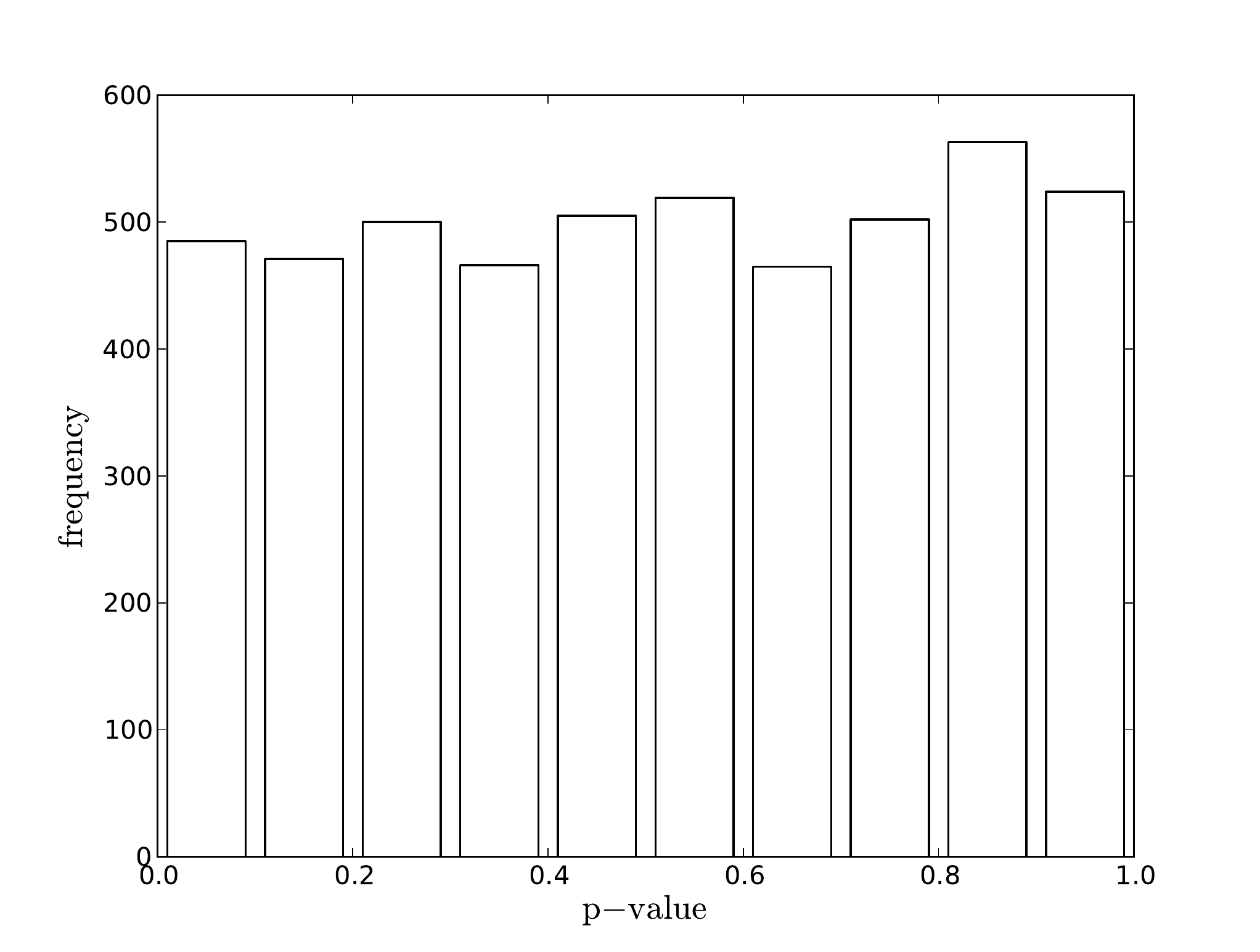}
        \end{center}
\caption{Distribution of p-values for 5000 simulations.}
\label{F8}
\end{figure}

Figure \ref{F8} shows a histogram of p-values generated by running the simulation 5000 times. We find that the distribution of p-values is reasonably uniform in the range $[0,1]$ indicating that indeed the increments $\Delta B'/\sqrt{\Delta t}$ belong to a normal distribution.

We conclude that the collapse data (the dots in Fig \ref{F7}) is distributed as though it had been generated by the backward in time collapse dynamics.

\section{Equilibrium and boundary conditions}
\label{equil}
The two examples considered so far consider specific models of objective collapse, exhibiting properties that look like an inherent time asymmetry (collapse events affecting the wave function after, but not before, the event; and positive correlation between the diffusion in mean position and momentum).  Our third example looks at a more generic feature of collapse models: non-conservation of mean energy.  This appears to give a directly observable time asymmetry.  In doing so, we will shed light upon the role of initial and final boundary conditions in causing time asymmetric behaviour.

It is well known (see, e.g.~\cite{GRW}) that one of the physical effects predicted by collapse models is a gradual increase in energy.  A monotonic mean energy increase can be seen to follow simply from Eq.(\ref{p}). The diffusion in momentum of an individual particle is such that momentum is just as likely to go up by some amount as it is to go down by the same amount. However, since energy is a convex function of momentum, $E \sim p^2$, the energy increases on average.  It is important to understand how this can be the case with a time symmetric picture since this increase in mean energy only appears in one time direction.

Consider a set of free localised particles, initially each with $p=0$ at time $t=0$. The total energy is zero. After a period of diffusion, the particles will have a Gaussian distribution of momenta centred on zero, but with a non-zero total energy.  Viewed as a process backward in time, the energetic particles would appear to simultaneously tend towards zero momentum and energy.  Even though we have shown that the collapse events are compatible with the Born rule applied to a backwards evolving wave function, the backwards in time evolution appears unlikely.  For example, we would find that the collapses undergone by particles with positive momentum would tend to cause them to reduce momentum, and the collapses undergone by particles with negative momentum would tend to cause them to increase momentum. By conditioning on the sign of the momentum, a reverse time observer would appear to see a conspiracy, that leads to the mean energy decreasing.  By observing the change in mean energy over time, the direction of time could be deduced.

The conspiratorial behaviour for the reverse time picture comes about as a result, not of an inherent asymmetry in the dynamics, but as a result of the time asymmetric use of boundary conditions.  From the forward in time picture, we have imposed the initial condition at time $t=0$ that $p=0$, and then tracked the behaviour for times $t>0$.  The evolution of the system then proceeds according to the Born rule, with no future boundary condition imposed.  Eventually this will result in a diffuse nearly uniform distribution.

Viewed in the reverse direction, we start with a diffuse distribution, evolving apparently according to the Born rule.  We might expect this distribution to stay diffuse (or possibly spread out more if it is not in equilibrium).  However, the behaviour we are tracking has a special feature that makes this impossible: these are pre-selected trajectories that must have come from localised $p=0$ states at time $t=0$.  This condition on the trajectories introduces a pre-selective bias in the statistics of the reverse time collapse events, and leads to the apparent conspiratorial behaviour.

To understand this more clearly, we will now address a well known problem for any stochastic theory \cite{UFF}: such a theory cannot exhibit time independent behaviour (such as that required by the Born rule) in both temporal directions, unless the probability distribution is in equilibrium.

The argument goes as follows: suppose a system follows a stochastic rule $R_{t_1|t_0}(S_j|S_i)$ for how it changes state from one time to another.  At time $t_0$, the probability for state $S_i$ is $P_{t_0}(S_i)$.  At some later time $t_1>t_0$, the stochastic evolution leads to the probability for state $S_j$ of
\begin{align}
    P_{t_1}(S_j)=\sum_i R_{t_1|t_0}(S_j|S_i)P_{t_0}(S_i).
\end{align}
From Bayes' theorem, the system being in state $S_j$ at time $t_1$ can be used to make retrodictions about the possible state at time $t_0$:
\begin{align}
	P_{t_0|t_1}(S_i | S_j) = R_{t_1|t_0}(S_j|S_i)\frac{P_{t_0}(S_i)}{P_{t_1}(S_j)}.
\end{align}

If the future directed stochastic evolution $R_{t_1|t_0}(S_j|S_i)$ is time independent (as we would expect from the Born rule), then adding any constant shift in time of $\tau$ will not affect the transition probabilities:
\begin{align}
R_{t_1+\tau|t_0+\tau}(S_j|S_i)=R_{t_1|t_0}(S_j|S_i).
\end{align}
However, imposing the same condition on the retrodictive probabilities
\begin{align}
	P_{t_0+\tau|t_1+\tau}(S_i | S_j) = P_{t_0|t_1}(S_i | S_j)
\end{align}
gives
\begin{align}
	P_{t_0|t_1}(S_i | S_j) = R_{t_1|t_0}(S_j|S_i)\frac{P_{t_0+\tau}(S_i)}{P_{t_1+\tau}(S_j)}.
\end{align}
which can only hold if
\begin{align}
	\frac{P_{t_0}(S_i)}{P_{t_0+\tau}(S_i)}=\frac{P_{t_1}(S_j)}{P_{t_1+\tau}(S_j)}=f(\tau)
\end{align}
Normalisation of probabilities yields $f(\tau)=1$. So the time independence condition cannot also hold in the backward direction, unless the system is in equilibrium with $P_{t_0}(S_i)=P_{t_0+\tau}(S_i)=P_E(S_i)$.  This would give a backwards in time retrodictive rule:
\begin{align}\label{eq:retro}
	P_{t_0|t_1}(S_i | S_j) = R_{t_1|t_0}(S_j|S_i)\frac{P_E(S_i)}{P_E(S_j)}.
\end{align}
which is time independant.

We argue that this does not imply any sort of time asymmetry in the dynamics.  Instead it is a result of the time asymmetric use of pre-selection statistics.  Let us instead start with the equilibrium distribution $P_E(S_i)$, and then impose a pre-selection at time $t=0$, that the system is in $S_0$.  For times $t_f>0$, we have, as before, the forward in time predictions $R_{t_f|0}(S_j|S_0)$.  If we now try to retrodict from the occurrence of $S_j$ at time $t_f$, to some intermediate time $0< t_1 <t_f$, we get:
\begin{align}
	P_{t_1|t_f,0}(S_i | S_j,S_0) = \frac{R_{t_f|t_1}(S_j|S_i)R_{t_1|0}(S_i|S_0)}{\sum_{i^\prime}R_{t_f|t_1}(S_j|S_{i^\prime})R_{t_1|0}(S_{i^\prime}|S_0)}
\end{align}
In general, this will not lead to the reverse time retrodictive rule Eq.(\ref{eq:retro}), nor will it be time independent. Pre-selecting the state $S_0$ at time $t=0$, biases the statistics of the retrodictive inferences in the time period $0< t_1 <t_f$, as they \textit{must} lead towards the state $S_0$ as $t_1 \rightarrow 0$.

However, exactly the same is true if we look at the forward in time statistics in the period $t_p < t_{-1} <0$.  First, if we impose the condition at time $t=0$, that the system is in $S_0$, then at times $t_p<0$ we have the retrodictive inference:
\begin{align}
	P_{t_p|0}(S_j | S_0) = R_{0|t_p}(S_0|S_j)\frac{P_E(S_j)}{P_E(S_0)}.
\end{align}
This leads to a time independent rule for $t_p<0$.  By contrast, if we try to make predictive inferences in the period $t_p < t_{-1} <0$, based on the occurrence of state $S_j$ at time $t_p$, we get
\begin{align}
	P_{t_{-1}|t_p,0}(S_i | S_j,S_0) = \frac{R_{0|t_{-1}}(0|S_i)R_{t_{-1}|t_p}(S_i|S_j)}{\sum_{i^\prime}R_{0|t_{-1}}(0|S_{i^\prime})R_{t_{-1}|t_p}(S_{i^\prime}|S_j)}
\end{align}
Again, in general, this will not lead to the rule $R_{t_{-1}|t_p}(S_i|S_j)$ nor will it be time independent.  Post-selecting the state at time $t=0$, biases the forward in time statistics in the period $t_p < t_{-1} <0$, as they \textit{must} lead towards the state $S_0$ as $t_{-1} \rightarrow 0$.

There should be nothing terribly mysterious about the fact that an apparently time independent, future directed stochastic evolution law $R_{t_1|t_0}(S_j|S_i)$ still does not yield time independent statistics when a post-selection is imposed at some time $t>t_1$.  The post-selection biases the statistics.  Equally so, therefore, there is nothing terribly mysterious that, when we start with a pre-selected ensemble, and look at the reverse time direction, the pre-selection introduces a bias in the retrodictive inferences.  The apparent time asymmetry associated with the bias in the statistics is a result of the pre-selection, rather than anything fundamental to the dynamics.

We may now return to the problem of non-conservation of energy in collapse models.   By pre-selecting a set of free localised particles, with $p=0$ at time $t=0$, as an initial condition, we are biasing the sample for the reverse time statistics when $t>0$.  An analogous situation can easily be constructed by starting with a wide and uniform distribution of initial momenta and post-selecting, at some time $t_f>t$, only those trajectories which end up with momentum equal to zero at $t=t_f$. The biases will now be present in the forward in time direction.  Post-selected particles with positive momentum would tend experience collapses which cause them to reduce momentum, and those with negative momentum would tend to increase.

Ideally, if the system has an equilibrium distribution, then to remove any biases we should start with samples whose distributions are approximately in equilibrium (see Ref.~\cite{BAC}).  In the case of the QMUPL model, the momentum is not bounded, and an equilibrium distribution of momentum will have an unbounded mean energy, even if the mean momentum is finite.  Any selection at $t=0$, with a finite mean energy, will be followed in the forward time direction by a diffusion towards equilibrium that increases the mean energy for $t>0$.  However, when looking at times $t<0$, the selection at $t=0$ is a post-selection.  The mean energy will be seen to be \textit{decreasing} in time, converging on the value fixed by the post-selection.  Once again, we see that the apparent asymmetry results from the use of initial or final boundary conditions, and not from any inherent asymmetry in the dynamics.  A time symmetric interpretation of the stochastic dynamical law is therefore not necessarily at odds with an observation of energy increase due to collapse.
\section{Summary}
\label{sum}

We have shown that collapse models can be viewed in a way in which they exhibit a time reversal symmetry. This is perhaps surprising given the apparent time-directedness of a collapse of the wave function. The key is to treat the collapse centres or flashes as the fundamental stuff of the theory and the wave function as part of the dynamical law used to determine where the next flash will occur. Indeed we have argued that for a given set of flashes there are two equivalent pictures of a collapsing wave function: one going forward in time and one going backward in time, each collapsing according to the Born rule.

This idea brings a new perspective to the problems of quantum theory. For example, in the usual picture of measurement the preceding quantum state is replaced by an eigenvector of the observable which is measured. Consider a particle passing through two misaligned polarisers. After passing through the first polariser the particle collapses to the first polarisation state. It stays in this state for the duration of its trajectory before it passes through the second polariser. In the backward in time picture the particle has the polarisation state of the second polariser at those points in time when it is situated between the two measurements. The resolution is to understand the wave function as an object which is part of the dynamical laws rather than part of the ontology. If we insist on understanding events in terms of the wave function then we must recognise that the wave function could be just as well determined by its future interactions as by its past \cite{PRI}.

\section*{Acknowledgements}
We would like to thank Sara Geneletti and Phil Pearle for useful comments. This work was funded by the Templeton World Charity Foundation.

\end{document}